\documentclass[aps,twocolumn,floatfix,showpacs,intlimits]{revtex4}
\usepackage{graphicx,bm}
\usepackage{amsmath}
\usepackage{amssymb}
\usepackage{bbm}
\begin{document}

\newcommand{\rb}{{\bf r}}
\newcommand{\Rb}{{\bf R}}
\newcommand{\xb}{{\bf x}}
\newcommand{\yb}{{\bf y}}
\newcommand{\zb}{{\bf z}}
\newcommand{\kb}{{\bf k}}
\newcommand{\Kb}{{\bf K}}
\newcommand{\kbp}{{{\bf k}^\prime}}
\newcommand{\Kbp}{{{\bf K}^\prime}}
\newcommand{\dk}{\frac{d^3k}{(2\pi)^3}\,}
\newcommand{\dK}{\frac{d^3K}{(2\pi)^3}\,}
\newcommand{\dkp}{\frac{d^3k^\prime}{(2\pi)^3}\,}
\newcommand{\dKp}{\frac{d^3K^\prime}{(2\pi)^3}\,}
\newcommand{\eb}[1]{{\bf \hat e_{#1}}}

\title{Ultracold bosons in the vicinity of a narrow resonance:\\ 
shallow dimer and recombination}
\author{Ludovic Pricoupenko$^{(1)}$ and Mattia Jona Lasinio$^{(2)}$}
\affiliation
{$^{(1)}$Laboratoire de Physique Th\'{e}orique de la Mati\`{e}re Condens\'{e}e,
Universit\'{e} Pierre et Marie Curie, case courier 121, 4 place Jussieu, 75252
Paris Cedex 05, France.
$^{(2)}$ Institut f\"{u}r Theoretische Physik , Leibniz Universit\"{a}t Hannover, Appelstr. 2, D-30167, Hannover, Germany\\
}
\date{\today}
\begin{abstract}
The different resonant regimes that can be achieved by using a magnetic Feshbach resonance are analyzed with a separable two-channel model. Emphasis is put on the case of narrow resonances in a region of intermediate detuning where a shallow dimer exists and an approximate law including the background scattering length for the three-body recombination rate is derived.
\end{abstract}
%
%03.65.Ge Solutions of wave equations: bound states
%03.65.Nk Scattering theory
%04.20.Cv Fundamental problems and general formalism
%05.30.Jp Boson systems (for static and dynamic properties of Bose-Einstein condensates, see 03.75.Hh and 03.75.Kk)
%31.15.ac Few-body systems atomic systems, calculations for, 
%32.80.Pj Optical cooling of atoms; trapping
%34.10.+x General theories and models of atomic and molecular collisions and interactions 
%	(including statistical theories, transition state, stochastic and trajectory models, etc.) 
%34.50.-s Scattering of atoms and molecules
%
\maketitle

\section{Introduction}

A possible way to prepare ultra-cold atoms in strongly correlated states is to use magnetic Feshbach resonances \cite{Fes58,Ino98}.  Using this techniques in Bose gases a breakthrough has been achieved with the observation of three-body Efimov states \cite{Kra06,Kno09,Zac09} and also four-body resonances tightly linked to the Efimov effect \cite{Fer09,Pol09}. However, a stable many-body  state has never been observed for bosons at resonance as a consequence of a large losses into deeply bound states \cite{Ste99}. In all these experiments, atoms are neutral and the pairwise interaction between them is characterized by a van der Waals tail. In the vicinity of a magnetic Feshbach resonance, the scattering length (denoted by $a$) can be tuned (formally from $-\infty$ to $\infty$)  \emph{via} an external magnetic field of amplitude denoted by ${\mathcal B}$.  In the vicinity of the resonance it can be parameterized as \cite{Moe95}:
\begin{equation}
a= a_{\rm bg} \left(1 - \frac{\Delta \mathcal B}{\mathcal B - \mathcal B_0} \right) ,
\label{eq:a}
\end{equation}
where  ${\mathcal B}_0$ is the position of the resonance, $\Delta {\mathcal B}$ is the width of the resonance such that $a=0$ for ${B={\mathcal B}_0+\Delta {\mathcal B}}$ and ${a_{\rm bg}}$ is the background scattering length, \emph{i.e.} the off-resonance $s$-wave scattering length.  Away from resonances  the two-body  scattering length is of the order of the van der Waals range \cite{Gri93,Dal98}:
\begin{equation}
R_{\rm vdW} =\frac{1}{2} \left(\frac{mC_6}{\hbar^2}\right)^{1/4} ,
\label{eq:RvdW}
\end{equation}
where $m$ is the mass of one atom and $C_6$ is the London's constant. In the resonant regime, this length is small as compared to the absolute value of the scattering length ${R_{\rm vdW}\ll |a|}$. The van der Waals length gives the short distance scale for interatomic forces and also permits to define a low energy regime corresponding to energies $E$ such that ${|E| \ll (\hbar^2/mR_{\rm vdW}^2)}$.

The precise description of the interaction between ultracold atoms needs multi-channel computations. In this paper we use instead a separable two-channel model which describes the Feshbach mechanism and contains the relevant energy scales in the vicinity of the resonance while short range properties are taken into account only qualitatively. It is a simplified version of the model of Refs.~\cite{Lee07,Chi10} and it was already introduced in the context of the two spin-component interacting Fermi gas and for the study of the three- and four-body problems in Refs.~\cite{Wer09,Jon10,Mor11a}. A similar version of the model has been also used in the context of Efimov physics for three spin-component fermions \cite{Nai10}. It has also the same structure as the one introduced in the study of three interacting polarized fermions in Ref.~\cite{Jon08}. In Refs.~\cite{Jon10,Mor11a} it has been shown that this separable two-channel model allows one to study the few-body problem by solving equations of the same order of difficulty than in the zero range approximation. The model takes into account the direct interaction between atoms and is thus able to describe the background scattering and the finite width of the resonance. In this paper, we use this feature in order to analyze the respective contributions of the background scattering length, of the width of the resonance and of the finite potential range on the properties of the shallow dimers in the different regimes encountered in experiments using the  magnetic Feshbach resonance techniques: broad and narrow resonances and also the possible neighborhood  of shape resonance. Deviations from universal laws valid near the threshold of resonance are obtained in these three regimes. We use the example of a broad resonance in potassium where the spectrum has been computed in a microscopic collisional approach in Ref.~\cite{Der07} to show that the basic description of the short range interaction with a single parameter is enough for a quantitative description of low energy properties. The case of narrow resonances, is particularly interesting because these deviations can be evaluated accurately without explicitly taking into account short range physics. For this type of resonances, we define an intermediate regime where the binding energy of the dimer is an affine law of the external magnetic field. Finally, we show that the two-channel model permits to study narrow resonances for a larger interval of magnetic detuning as compared to the effective range approach \cite{Pet04b}. Finally, we find an approximate law for the recombination constant which takes into account the background scattering length $a_{\rm bg}$ and also generalizes the analogous law obtained in the framework of the effective range approximation.

The paper is decomposed as follows. In section \ref{sec:model} we define the two-channel separable model used in this paper.  In section \ref{sec:two-body} we analyze the basic properties of the shallow dimers in the different regime of the  the magnetic Feshbach resonance. In section \ref{sec:three-body} we derive the integral equation which permits to solve the three-body problem. We then consider the recombination rate issue for three identical bosons and apply the formalism to the case of the intermediate regime of narrow resonances defined in this paper. 

\section{Two-channel separable model}

\label{sec:model}

We consider neutral and spinless bosonic atoms. They belong to the open channel of the model characterized by a continuum of delocalized states above zero energy. The model supports a closed channel characterized by a discrete and structureless bosonic molecular state. In our terminology, molecules in the closed channel are distinct from the dimers which are the bound states of the full Hamiltonian. The Feshbach resonance mechanism is encapsulated by a coherent coupling between atomic pairs of the open channel and the molecular state. The model contains a direct interaction term between atoms in the open channel which permits to take into account the background \emph{i.e.} off-resonance scattering length (denoted by ${a_{\rm bg}}$).

We use the second quantized form, where the operator ${a_\kb}$ annihilates one atomic boson of mass $m$ and wavevector $\kb$ in the open channel while ${b_\kb}$ annihilates one molecule of mass $2m$ and wavevector $\kb$ in the closed channel. Both operators ${a_\kb}$ and ${b_\kb}$ obey standard bosonic commutation rules:
\begin{equation}
[a_\kb,a_{\kbp}^\dagger]=(2\pi)^3 \delta(\kb-\kbp) \ , \  [b_\kb,b_{\kbp}^\dagger]=(2\pi)^3 \delta(\kb-\kbp), 
\label{eq:commut}
\end{equation}
where the factor ${(2\pi)^3}$ in front of the $\delta$-distribution corresponds to the choice ${\langle \rb|\kb \rangle=\exp({i \kb \cdot \rb})}$ for the plane  wave. Any other commutator vanishes. In what follows, the kinetic energy of an atomic plane wave of wave vector ${\mathbf k}$ is
\begin{equation}
\epsilon_\kb = \frac{\hbar^2 k^2}{2m} .
\label{eq:epsilon_k}
\end{equation}
The Hamiltonian of the model can be written as the sum of three terms:
\begin{equation}
H = H_{\rm at} + H_{\rm mol} + H_{\rm at-mol} ,
\label{eq:H}
\end{equation}
where ${H_{\rm at}}$ is the Hamiltonian for bosons in the open channel, ${H_{\rm mol}}$ is the Hamiltonian for molecules and ${H_{\rm at-mol}}$ couples any atomic pair with the molecular state. The three terms in Eq.~\eqref{eq:H} are defined as follows:
\begin{itemize}

\item The Hamiltonian ${H_{\rm at}}$ describes atoms:
\begin{multline}
H_{\rm at} =\int \dk \epsilon_{\mathbf{k}}  a^\dagger_\kb a_\kb + \frac{g_0}{2} \int \frac{d^3K d^3k d^3k^\prime} {(2\pi)^9}  \chi^*_\kb \chi_\kbp\\ 
a^\dagger_{\frac{\Kb}{2} -\kbp} a^\dagger_{\frac{\Kb}{2}+\kbp} a_{\frac{\Kb}{2}+\kb} a_{\frac{\Kb}{2}-\kb} .
\label{eq:Hat}
\end{multline}
For simplicity, the pairwise interaction among atoms has been chosen separable: it is characterized by a coupling constant ${g_0}$ and a normalized function ${\chi_{\mathbf k}}$ with a Gaussian shape and a cut-off length denoted by $b$:
\begin{equation}
\chi_\kb=\exp(-k^2b^2/2) .
\label{eq:chi_k}
\end{equation}
The parameter $b$ mimics the potential range of the true interaction between neutral atoms and is thus of the order of the van der Waals range ${R_{\rm vdW}}$. The particular choice for the shape of the  function ${\chi_\kb}$ is not essential: it allows to perform some analytical simplifications and gives a qualitative description of the short range physics. The model Hamiltonian ${H_{\rm at}}$ allows one to describe  two-body scattering far from the Feshbach resonance: the background scattering length of the model is a function of ${g_0}$ and $b$. 

\item Molecules in the closed channel are described by $H_{\rm mol}$:
\begin{equation}
H_{\rm mol}=\int \dK \left( \frac{ \epsilon_{\mathbf K}}{2} +E_{\rm mol} \right) b^\dagger_\Kb b_\Kb ,
\end{equation} 
where $E_{\rm mol}$ is the internal energy of a molecule defined with respect to the zero energy of the open channel. In the magnetic Feshbach resonances,   $E_{\rm mol}$ is tuned by use of the external magnetic field $\mathcal B$.  We make the reasonable assumption that in the vicinity of the resonance, it can be approximated by an affine law:
\begin{equation}
E_{\rm mol}=\delta\mu ({\mathcal B}-{\mathcal B}_0^{\rm cl}) .
\label{eq:Emol} 
\end{equation}
In Eq.(\ref{eq:Emol}), ${\mathcal B}_0^{\rm cl}$ is the magnetic field at which the  molecular state energy crosses  the threshold of the open channel continuum and $\delta \mu$ denotes the difference between the magnetic moments for an atomic pair in the open channel and the molecular state in the closed channel.

\item The last term in Eq.~\eqref{eq:H} couples the two channels and thus models the Feshbach resonance mechanism:
\begin{equation} 
H_{\rm at-mol} = \Lambda \int 
\frac{d^3Kd^3k}{(2\pi)^6}  \, \chi^*_{\bf k} \, b^\dagger_{\bf K} a_{\frac{\Kb }{2}-\kb} a_{\frac{\Kb}{2}+\kb} + {\rm h. c.}  .
\label{eq:at-mol}
\end{equation}
In Eq.~\eqref{eq:at-mol}, the coupling function is generic: it is characterized by a constant $\Lambda$ (chosen here to be real) and the same cut-off function $\chi_\kb$ as in Eq.~\eqref{eq:chi_k}. When applied in the context of the few-body problem, this particular choice greatly simplifies the original model of Ref.~\cite{Lee07}: it permits to obtain single closed equations for the three- and four-body problems \cite{Jon08,Mor11a} similar to the ones obtained in the zero-range potential approach \cite{Pri11}. 

\end{itemize}

\section{Two-body properties}

\label{sec:two-body}

\subsection{Scattering amplitude}

The basic object for the study of two-body properties is the scattering amplitude. Thus, in this subsection we derive the two-body scattering amplitude of the two-channel model to make the link between two-body properties which may be measured experimentally ($a$,${a_{\rm bg}}$,${\Delta \mathcal B}$, ${E_{\rm dim}}$) and the four  parameters of the model (${g_0}$, ${\Lambda}$, ${E_{\rm mol}}$,$b$).

The most general state for the two-body system in the center of mass frame is a coherent superposition of two atoms in the open channel and of one molecule at rest in the closed channel:
\begin{equation}
\label{eq:2b-psi-ansatz}
|\Psi \rangle = \int \dk  A_{\kb} a^\dagger_{\kb} a^\dagger_{-\kb} | 0 \rangle +
\beta b^\dagger_{\bf 0} |0 \rangle .
\end{equation}
$A(\kb)$ is thus the wavefunction for a pair of atoms. The equations verified by the amplitudes  $A(\kb)$ and $\beta$ in Eq.~\eqref{eq:2b-psi-ansatz} are obtained from the projection onto the atomic and molecular subspaces of the stationary Schr\"{o}dinger's  equation at energy $E$: ${(E-H)|\Psi\rangle=0}$:
\begin{eqnarray} 
\label{eq:schrod-A}
&&\left( E - 2 \epsilon_{\mathbf k} \right) A_{\kb} = \Lambda \chi_\kb \beta + g_0 \chi_\kb \gamma \\ 
&&(E-E_{\rm mol}) \beta = 2 \Lambda  \gamma  ,
\label{eq:schrod-beta}
\end{eqnarray}
where ${\gamma = \int \dkp \chi^*_\kbp A_\kbp}$. From Eqs.~(\ref{eq:schrod-A},\ref{eq:schrod-beta}) we eliminate $\gamma$ and obtain:
\begin{equation}
\left(E- 2 \epsilon_{\mathbf{k}} \right)A_\kb- \frac{g_0  \chi_\kb}{2
\Lambda}    \left( E-E_{\rm mol}+\frac{2\Lambda^2}{g_0}\right) \beta  = 0 .
\label{eq:intermediaire}
\end{equation}
For a scattering process, the atomic wavefunction is characterized by an incoming plane wave of momentum ${\kb_0}$ and an outgoing spherical wave:
\begin{equation}
A_\kb = (2\pi)^3\delta(\kb-\kb_0) + \frac{g_0 \chi_\kb}{2 \Lambda}
\frac{E- E_{\rm mol} + \frac{2\Lambda^2}{g_0}}{E-2 \epsilon_{\mathbf{k}}+i0^+}
\beta,
\label{eq:A(k)}
\end{equation}
where we use the standard prescription ${E \to E+ i0^+}$ to ensure that the scattered wave is indeed outgoing. General scattering theory  relates the scattering state ${|\Psi \rangle}$ with the incoming state  ${|\Psi^{\rm (0)} \rangle}$ by ${|\Psi \rangle = (1+G_0 T) |\Psi^{\rm (0)} \rangle}$,  where $T$ denotes the transition operator and $G_0$ is the resolvent of the non-interacting Hamiltonian \cite{Tay72}. From this identity and  from Eq.~\eqref{eq:A(k)}, the half on-shell transition matrix is:
\begin{equation}
\langle \kb | T(E+i0^+) | \kb_0 \rangle =  \frac{g_0 \chi_\kb}{2 \Lambda} \left( E- E_{\rm mol} + \frac{2\Lambda^2}{g_0} \right)\beta.
\end{equation}
The scattering amplitude $f(E)$ is derived from the on-shell transition matrix ($k=k_0$) using the relation:
\begin{equation}
f(E)=-\frac{m}{4\pi \hbar^2}  \langle \kb | T(E+i0^+) | \kb_0 \rangle .
\end{equation}
This way, $f(E)$ can be expressed as:
\begin{multline}
\frac{m |\chi_{\mathbf k_0}|^2}{4\pi\hbar^2f(E)}= \int \dk  \frac{|\chi_{\mathbf
k}|^2} {E-2 \epsilon_{\mathbf{k}}+i0^+}  \\
- \frac{1}{g_0} + \frac{2 \Lambda^2}{g_0^2(E-E_{\rm mol}+\frac{2\Lambda^2}{g_0})} .
\label{eq:scatter-amp}
\end{multline}
Equation~\eqref{eq:scatter-amp} shows that in this model scattering only occurs in the $s$-wave channel which is a sufficient description of binary collisions  in the regime of temperature and densities where ultracold gases are studied. In the absence of the inter-channel coupling (${\Lambda=0}$) {\emph i.e.} in the off-resonant regime, the scattering amplitude at zero energy is by definition equals to the opposite of the background scattering length $a_{\rm bg}$ and from Eq.~\eqref{eq:scatter-amp}, one finds:
\begin{equation}
a_{\rm bg} = \frac{bg_0\sqrt{\pi}}{g_0-g_0^{\rm c}} \quad  \mbox{with}\quad 
g_0^{\rm c} = - \frac{4\pi^{3/2}\hbar^2 b}{m} .
\label{eq:abg}
\end{equation}
The system is in the neighborhood of a shape resonance when ${|a_{\rm bg}| \gg b}$ and in this model this corresponds to the situation where ${g_0 \sim  g_0^{\rm c}}$. Away from a shape resonance, ${|a_{\rm bg}|}$ is of the order of the potential range $b$.

The general calculation of the scattering amplitude at finite energy in Eq.~\eqref{eq:scatter-amp} is easy to perform in the domain of negative energy  ${E=\hbar^2 k_0^2/m<0}$. Keeping in mind the standard analytical continuation  ${k_0 = iq}$ with ${q > 0}$, one obtains:
\begin{equation} 
\frac{1}{f(E)}= q \mbox{erfc}(q b) - \frac{e^{-q^2 b^2}}{a_{\rm bg}} 
\left( 1 - \frac{\frac{8 \pi \hbar^2 \Lambda^2 a_{\rm bg}}{m g_0^2}}{E-E_{\rm mol}+\frac{2\Lambda^2}{g_0}}\right),
\label{eq:scat-amplitude}
\end{equation}
where erfc() is the complementary error function which equals unity at zero. The zero energy limit of Eq.~\eqref{eq:scat-amplitude} is equal to the opposite of the inverse scattering length that can be {\it exactly} identified with the standard phenomenological expression of Eq.~\eqref{eq:a}. This provides the expression of the width of the resonance as a function of the parameter of the model:
\begin{equation}
\Delta {\mathcal B} =  \frac{8 \pi \hbar^2 \Lambda^2 a_{\rm bg}}{m g_0^2 \delta \mu},
\label{eq:DeltaB}
\end{equation}
and also the magnetic detuning from the center of the resonance:
\begin{equation}
{{\mathcal B}-{\mathcal B}_0} = \frac{E_{\rm mol}}{\delta \mu} - \frac{2\Lambda^2}{g_0 \delta \mu }  + {\Delta \mathcal B} .
\label{eq:detuning}
\end{equation}
In what follows, ${\delta \mu ({\mathcal B}-{\mathcal B}_0)}$ is called the energy detuning. Equation \eqref{eq:DeltaB} shows that ${\delta \mu \Delta \mathcal B}$ has the same sign as the background scattering length $a_{\rm bg}$. We have checked that this property is indeed verified for the various resonances reported in Ref.~\cite{Chi10}. As a consequence of the interaction between the open and the closed channel the resonance is not located at ${\mathcal B}_0^{\rm cl}$ and from Eqs.~(\ref{eq:Emol},\ref{eq:detuning}) the resonance shift in the magnetic field is:
\begin{equation}
{\mathcal B}_0^{\rm cl}-{\mathcal B}_0 = \frac{\Delta {\mathcal B}}{1-b\sqrt{\pi}/a_{\rm bg}} .
\label{eq:shift}
\end{equation}
One can notice that nearby a shape resonance where ${|a_{\rm bg}| \gg b}$  the resonance shift appears as independent of the short range details of the interatomic forces and equals $-\Delta B$. This is the case of the the Feshbach resonance in cesium at {${-12}$~G} where one finds ${({\mathcal B}_0^{\rm cl}-{\mathcal B}_0 )/\Delta \mathcal B \simeq -1.12}$ if one takes the reasonable value ${b = R_{\rm vdW}}$ for the potential range.

\subsection{Dimers}

\label{sec:dimers}

\subsubsection{Binding energies}

Using the results of the previous section, we now discuss the properties of dimers \emph{i.e.} the two-body bound states of the full Hamiltonian in Eq.~\eqref{eq:H}. Their binding energies (denoted by ${E_{\rm dim}= \frac{\hbar^2 q_{\rm dim}^2}{m} >0}$) are given by the poles of the scattering amplitude $f(E)$ at negative energy (${E=-E_{\rm dim}<0}$) \cite{Lan99}. From  Eq.~\eqref{eq:scat-amplitude} this leads to the following equation:
\begin{multline}
\frac{q_{\rm dim} a_{\rm bg}\,e^{q_{\rm dim}^2 b^2}\mbox{erfc}(q_{\rm dim} b) }{q_{\rm dim} a_{\rm bg}\,e^{q_{\rm dim}^2 b^2}\mbox{erfc}(q_{\rm dim} b)- 1} 
\\ 
- \frac{\hbar^2q_{\rm dim}^2}{m {\delta \mu \Delta \mathcal B}} = \frac{\mathcal B - \mathcal B_0}{\Delta \mathcal B} .
\label{eq:poles}
\end{multline}
In the limit where $E$ and ${\delta \mu (\mathcal B - \mathcal B_0)} $ have both large and negative values, the leading term in the expansion of the left hand side gives the solution ${E=E_{\rm mol}= -E_{\rm dim}}$: \emph{i.e.}, far from resonance the dimer coincides with the molecular state and is essentially in the closed channel. At finite detunings, one may distinguish two qualitatively different regimes in Eq.~\eqref{eq:poles}:
\begin{itemize}
\item[(i)] For ${a_{\rm bg} < b \sqrt{\pi}}$, the left hand side of Eq.~\eqref{eq:poles} is negative and is a monotonically decreasing function of ${q_{\rm dim}}$.
Hence, there is at most one possible solution which only exists for a negative energy detuning ${\delta \mu (\mathcal B - \mathcal B_0)}$. The dimer results from the inter-channel coupling and is denoted below as the Feshbach dimer.
\item[(ii)] For ${a_{\rm bg} > b \sqrt{\pi}}$, the left hand side of  Eq.~\eqref{eq:poles} diverges at a given value denoted in the following by ${q^{\rm div}}$. It decreases from $0$ to $-\infty$ in the interval ${q_{\rm dim} \in [0,q^{\rm div}[}$ and from $+\infty$ to $-\infty$ in the interval ${q_{\rm dim} \in ]q^{\rm div},+\infty[}$. Consequently, there always exists a solution in the interval ${]q^{\rm div}, +\infty[}$. Away from the Feshbach resonance and for positive energy detunings [${{\delta \mu (\mathcal B - \mathcal B_0)}  \to +\infty}$], this solution gives the energy of the dimer associated with the direct coupling and we denote it as the background dimer. For negative energy detunings [${\delta \mu (\mathcal B - \mathcal B_0) <0}$] a second dimer exists: at threshold [\emph{i.e.} for ${{\delta \mu (\mathcal B - \mathcal B_0)}  =0^-}$] it corresponds to the Feshbach dimer and for decreasing values of the energy detuning there is an avoided crossing between the Feshbach and the background dimers.
\end{itemize}
Due to their important role and their universal character, we now focus on the low energy two-body properties of this model. We thus assume that the dimer's energy is  such that ${q_{\rm dim} b \ll 1}$. In order to obtain an approximation beyond the universal regime where ${q_{\rm dim}=1/a}$, we solve the equation:
\begin{equation}
\frac{E+{\delta \mu (\mathcal B_0 - \mathcal B + \Delta \mathcal B)}}{f(E)} = 0 ,
\end{equation}
by expanding expanding the left hand side up to second order in $q$ and find:
\begin{equation}
q_{\rm dim}  \simeq \frac{-a+\sqrt{a^2+4 R^\star(a-a_{\rm bg})-\frac{8ba}{\sqrt{\pi}}+4b^2}}{2R^\star(a-a_{\rm bg})-\frac{4ba}{\sqrt{\pi}}+2b^2} ,
\label{eq:q_LowE}
\end{equation}
where the parameter ${R^\star}$ denoted hereafter as the width radius is linked to the width of the resonance by \cite{Pet04b}:
\begin{equation}
R^\star =  \frac{\hbar^2}{m a_{\rm bg} {\delta \mu \Delta \mathcal B}}.
\label{eq:Rstar}
\end{equation}
We anticipate that the expression in Eq.~\eqref{eq:q_LowE} cannot be obtained from the effective range approach (even in the formal limit where ${b \to 0}$). In the limit of large and positive values of the scattering length, \emph{i.e.} at the threshold of appearance of the Feshbach dimer, Eq.~\eqref{eq:q_LowE} gives at the lowest order the universal and standard law supported also by the Bethe-Peierls zero-range model \cite{Bet35}:
\begin{equation}
q_{\rm dim} = \frac{1}{a} \quad ; \quad E_{\rm dim} = \frac{\hbar^2}{m a^2} .
\label{eq:Ebp}
\end{equation}
The interest of the two-channel model is to give quantitative corrections to this law. In the vicinity of the threshold where ${a}$ is much larger than the three length  ${b}$, ${R^\star}$ and ${|a_{\rm bg}|}$, the correction in ${1/a^2}$ on ${q_{\rm dim}}$ is given by:
\begin{equation}
q_{\rm dim} \simeq \frac{1}{a} \left(1-\frac{R^\star}{a}+\frac{2b}{\sqrt{\pi} a}\right) + \dots
%les termes en 1/a^3 ne sont pas consistants avec l'approximation faite sur qdim (on a négligé les termes à partir de q^3 et donc en 1/a^3)
\label{eq:qdim_dl}
\end{equation}
where terms of higher order are not consistent with the approximation made for Eq.~\eqref{eq:q_LowE}. A similar law was derived by using another two-channel model in Ref.~\cite{Chi10} (see Eq.~(53) in this last reference). We now distinguish different regimes where the deviation to the universal prediction are qualitatively different. 

\paragraph{Broad resonances--} For a broad resonance, ${R^\star}$ is of the same order of magnitude or smaller than $b$, meaning that the correction to the universal law depends on the details of the true interatomic potential and Eq.~\eqref{eq:Ebp} is valid in the regime of small detuning: 
\begin{equation}
\frac{|{\mathcal B}-{\mathcal B}_0|}{|\Delta {\mathcal B}|} \ll 1 .
\end{equation}

\paragraph{Narrow resonances--} The case of narrow resonances  defined by ${R^\star \gg |a_{\rm bg}|}$ has attracted some interest in the three-bosons problem \cite{Pet04b,Gog08,Pri10b} and we focus on the dimer spectrum given by the present realistic model. Near threshold for this type of resonance, the regime of validity of Eq.~\eqref{eq:Ebp} is very tiny:
\begin{equation}
\frac{|{\mathcal B}-{\mathcal B}_0|}{|\Delta {\mathcal B}|} \ll \frac{|a_{\rm bg}|}{R^\star} \ll 1.
\end{equation}
However for a narrow resonance, the first correction to Eq.~\eqref{eq:Ebp} given by Eq.~\eqref{eq:qdim_dl} has also an universal character: a feature already noticed in Refs.~\cite{Pet04b,Chi10}. A more interesting feature brought by the two-channel model concerns higher relative detuning ${|{\mathcal B}-{\mathcal B}_0|/|\Delta {\mathcal B}|}$, where there still exists a shallow dimer. Equation~\eqref{eq:q_LowE} permits to distinguish between the contributions of the background scattering length ${a_{\rm bg}}$ and of the potential range $b$. In the regime where ${R^\star \gg b \left|\frac{\mathcal B-\mathcal B_0}{\Delta \mathcal B}\right|}$,  terms involving $b$ are negligible and one can thus use the following approximate law:
\begin{equation}
q_{\rm dim}  = \frac{-a+\sqrt{a^2+4 R^\star(a-a_{\rm bg})}}{2R^\star(a-a_{\rm bg})} .
\label{eq:q_narrow}
\end{equation}
Moreover, in the intermediate regime where ${R^\star(a-a_{\rm bg}) \gg a^2}$  the binding energy of the Feshbach dimer is given by:
\begin{equation}
q_{\rm dim} \simeq \frac{1}{\sqrt{R^\star (a-a_{\rm bg})}} \quad ; \quad E_{\rm dim} \simeq \delta \mu ({\mathcal B}_0-{\mathcal B}) .
\label{eq:intermediate1}
\end{equation}
This law, derived in the limit of a large width radius ${R^\star}$, shows that for a negative background scattering length ${(a_{\rm bg}<0)}$ and  for a vanishing value of the scattering length ${(|a|\ll |a_{\rm bg}|)}$, there still exists a shallow dimer of binding energy ${E_{\rm dim}= - \delta \mu \Delta \mathcal B}$. In terms of magnetic field detuning, the  definition of the intermediate regime chosen here can be written as:
\begin{equation}
\frac{R^\star}{|a_{\rm bg}|} \gg \left| \frac{\Delta \mathcal B}{\mathcal B - \mathcal B_0} + \frac{\mathcal B - \mathcal B_0}{\Delta \mathcal B} \right| .
\label{eq:intermediate_new}
\end{equation}
It corresponds to a larger interval of detuning than the intermediate regime of Ref.~\cite{Pet04b} where it is defined as ${R^\star \gg |a|\gg |a_{\rm bg}|}$ and corresponds to  the inequalities: ${\frac{R^\star}{|a_{\rm bg}|} \gg \left| \frac{\Delta \mathcal B}{\mathcal B - \mathcal B_0} \right| \gg 1 }$. 

\begin{figure}[h]
\includegraphics[width=8cm]{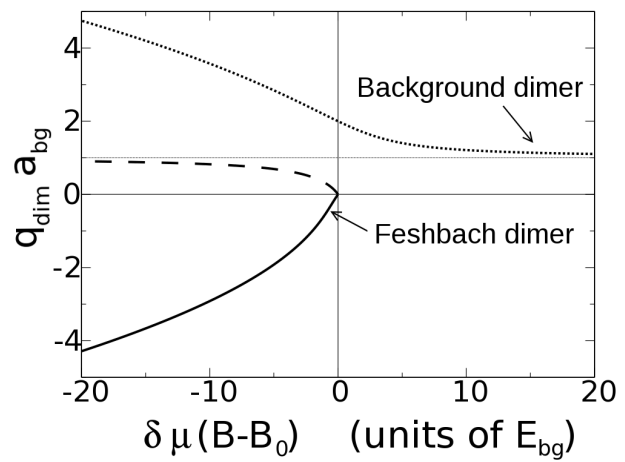}
\caption{Graphical solution of Eq.~\eqref{eq:poles} for ${q_{\rm dim}b \ll 1}$ and  the ratio ${|a_{\rm bg}|/R^\star=2}$. The product ${q_{\rm dim} a_{\rm bg}}$  is plotted as a function of the energy detuning ${\delta \mu (\mathcal B - \mathcal B_0)}$. In the regime of negative background scattering length $({a_{\rm bg}<0})$ only one shallow dimer exists (solid line). For positive scattering length ${(a_{\rm bg}<0)}$ there exists an avoided crossing between the Feshbach dimer and the background dimer respectively (dashed and dotted lines respectively).}
\label{fig:interplay}
\end{figure}

\paragraph{Vicinity of a shape resonance--} In the case where the system is in the vicinity of a shape resonance \emph{i.e.} ${|a_{\rm bg}|\gg b}$ and for a positive background scattering length, far from the Feshbach threshold the background dimer is shallow. In the regime where $a$ of the same order of magnitude than ${a_{\rm bg}}$, the expansion of Eq.~\eqref{eq:q_LowE} for large values of $a_{\rm bg}$ (for a fixed ratio ${a_{\rm bg}/a}$) gives the correction to the universal law ${q_{\rm dim} \sim 1/a_{\rm bg}}$: 
\begin{equation}
q_{\rm dim} \simeq \frac{1}{a} \left[ 1-\frac{R^\star}{a} (1-\frac{a_{\rm bg}}{a}) +\frac{2b}{\sqrt{\pi} a}\right] + \dots 
\label{eq:shape}
\end{equation}
The presence of the parameter $b$ in Eq.~\eqref{eq:shape} shows that this correction depends on the details of the real interatomic forces at short distances. For arbitrary magnetic detuning, the set of solutions of Eq.~\eqref{eq:poles} can be obtained by using a graphical reasoning. As an example, in Fig. (\ref{fig:interplay}) we have plotted the possible values of $q_{\rm dim} a_{\rm bg}$ in Eq.~\eqref{eq:poles}  as a function of the energy  detuning ${\delta \mu (\mathcal B - \mathcal B_0)} $ for a specific value of the ratio $|a_{\rm bg}|/R^\star$. For a negative  background scattering length (${a_{\rm bg} \ll -b }$) only Feshbach dimers exist and the corresponding branch is plotted in the negative region of the product ${q_{\rm dim} a_{\rm bg}}$ and of the energy detuning. For a positive background scattering length (${a_{\rm bg} \gg  b}$),  $q_{\rm dim} a_{\rm bg}$ is positive and there are two possible branches displayed in Fig.~(\ref{fig:interplay}). In this regime, the figure nicely illustrates the interplay between the Feshbach dimer and the background dimer \cite{Mar04}.
At the Feshbach threshold [${\delta \mu (\mathcal B-\mathcal B_0)}=0$], the binding energy of the dimer of finite energy is:
\begin{equation}
E_{\rm dim} \simeq \frac{\hbar^2}{4m a_{\rm bg}^2 } \left(1+ \sqrt{1+ 4 a_{\rm bg}/R^\star} \right)^2.
\end{equation}
Hence, as we consider the vicinity of a shape resonance where ${|a_{\rm bg}| \gg R_{\rm vdW}}$, this bound state is shallow only for a sufficiently narrow resonance. This shows the possible occurrence of a regime where two low energy dimers coexist. Interestingly, in the case of a narrow resonance in the vicinity of a shape resonance Eq.~\eqref{eq:q_narrow} is a low energy solution for all values of the magnetic detuning including the off-resonant regime where ${a}$ is of the order of ${a_{\rm bg}}$.

\subsubsection{Occupation probability in the closed channel}

The inter-channel coupling induces a non vanishing occupation probability of the dimer in the closed channel. In what follows, this probability is denoted by ${p_{\rm closed}}$. For a normalized bound state: 
\begin{equation}
|\beta|^2 + 2\int \dk |A_\kb|^2 = 1 
\end{equation}
and ${p_{{\rm closed}}=|\beta|^2}$. In the case of a shallow dimer such that $q_{\rm dim} b \ll 1$, this probability is independent of the short-range parameter $b$:
\begin{equation}
p_{\rm closed}=\frac{q_{\rm dim}}{q_{\rm dim}+\frac{1}{2R^\star} \left(\frac{E_{\rm dim}}{\delta \mu \Delta \mathcal B} +\frac{a}{a_{\rm bg} -a} \right)^2} .
\label{p-closed}
\end{equation}
Near the Feshbach threshold, for a vanishing and negative value of the energy detuning ${\delta \mu (\mathcal B - \mathcal B_0)}$, the Feshbach dimer has a vanishing occupation probability in the closed channel:
\begin{equation}
p_{\rm closed} \sim \frac{2 R^\star}{a} ,
\label{eq:pclosed-Feshbach}
\end{equation}
while in the intermediary regime of Eq.~\eqref{eq:intermediate_new}, this occupation probability is close to unity: 
\begin{equation}
p_{\rm closed} \sim 1- \frac{1}{2} \sqrt{\frac{a-a_{\rm bg}}{R^\star}} .
\label{eq:pclosed-intermediate}
\end{equation}
 
\subsection{Short range parameter of the model}

In standard broad resonances used in current experiments, for a detuning ${|{\mathcal B}-{\mathcal B}_0|}$ of few Gauss and even if ${E_{\rm dim} < E_{\rm vdW}}$,  the deviation of the dimer's binding energy from the universal threshold law of Eq.~\eqref{eq:Ebp} is non negligible. Interestingly, this property can be used as a way to set a reasonable value for the short range parameter $b$. In order to justify that $b$ is of the same order of magnitude than ${R_{\rm vdW}}$, we use an approximate formula which has been derived semi-classically for a one-channel model with a potential containing a van der Waals tail \cite{Gri93}:
\begin{equation}
E_{\rm dim} \simeq \frac{{\hbar}^2}{m(a-\overline{a})^2} \quad \mbox{where,} \quad \overline{a}= \frac{\Gamma(3/4)R_{\rm vdW}}{\sqrt{2}\Gamma(5/4)} .\label{eq:Gribakin}
\end{equation}
It has been shown that this approximation is relevant in the limit of large scattering length ${a\gg R_{\rm vdW}}$ and for a broad resonance, where the dimer near threshold is essentially in the open channel \cite{Koh06,Chi10}. Comparing this law to Eq.~\eqref{eq:qdim_dl} confirms our assumption on the parameter $b$. In the case of narrow resonances where $R^\star$ is large as compared to ${|a_{\rm bg}|}$, Eq.~\eqref{eq:q_LowE} shows that low energy two-body properties depend only sightly on the precise choice made for $b$ which is of the order of ${R_{\rm vdW}}$. 

\begin{figure}[hx]
\begin{center}
\includegraphics[width=8cm,clip]{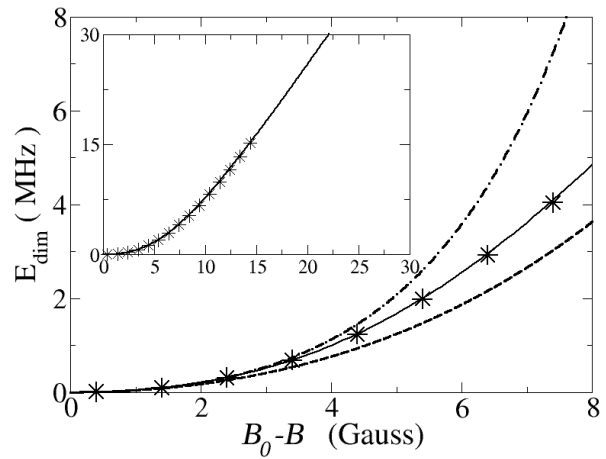}
\caption{Spectrum of the Feshbach dimer for the resonance at $402$~G in $^{39}$K. Stars: microscopic computations~\cite{Der07}; continuous line: two-channel model used in this paper where $b=1.2$~$R_{\rm vdW}$; dashed line: universal law ${E_{\rm dim}=\hbar^2/(ma^2)}$; dashed dotted line: semi-classical prediction from Eq.~\eqref{eq:Gribakin}. Inset: for large detunings, the binding energy tends to $-E_{\rm mol}$.}
\label{fig:compardimers1}
\end{center}
\end{figure}

For a precise determination of $b$, one may use data provided by detailed collisional models or experimental results when available. As an example, the spectrum for the resonance at $402$~G in $^{39}$K has been computed in Ref.~\cite{Der07} (see Fig.~4 in this last reference). It appears that the choice ${b=1.2 R_{\rm vdW}}$ permits to fit precisely the spectrum even for a relatively large detuning. We plotted in Fig.~\eqref{fig:compardimers1} the different predictions: the semi-classical law \eqref{eq:Gribakin}, the universal law \eqref{eq:Ebp}, the microscopic computation of Ref.~\cite{Der07} and the result given by the present two-channel model. 
In Ref.~\cite{Jon08}, we have also considered the resonance at -12~G for $^{133}$Cs where experimental results of spectroscopy were obtained \cite{Mar07}. We have shown also that the data can be reproduced by a specific value of the short range parameter which is close to the van der Waals range (${b=0.7}$~${R_{\rm vdW}}$). These studies provides strong confidences that for the various magnetic Feshbach resonances that can be encountered in experiments,  the model can reproduce accurately two-body properties for energies lesser than $E_{\rm vdW}$.

\subsection{Effective range approximation}

In the effective range approach, the inverse scattering amplitude  ${1/f(E)}$ is replaced by its approximation at the linear order in $E$:
\begin{equation}
\frac{1}{f(E)} = - \frac{1}{a} + q - \frac{ r_{\rm e} q^2}{2}  \quad \mbox{for} \quad E<0 .
\label{eq:effective_range}
\end{equation}
In Eq.~\eqref{eq:effective_range} ${r_{\rm e}}$ is the effective range and ${q=-ik}$ if ${E>0}$. In what follows, we compare the  predictions for the shallow dimer obtained from Eq.~\eqref{eq:effective_range} to the one  given by the two-channel model, we thus choose the value of the effective range obtained from Eq.~\eqref{eq:scat-amplitude}:
\begin{equation}
r_{\rm e} = \frac{4b}{\sqrt{\pi}} - \frac{2b^2}{a} - 2 R^\star \left( 1- \frac{a_{\rm bg}}{a} \right)^2 .
\label{eq:re}
\end{equation}
At resonance the effective range is ${r_{\rm e}^{\rm res} = \frac{4b}{\sqrt{\pi}} -2 R^\star}$, thus ${r_{\rm e}^{\rm res}}$ has a positive upper-bound of the order of the short range length $b$ but depending on the parameters of the model, it can in principle assume any negative value. For a broad Feshbach resonance the effective range is of the order of the van der Waals range while for a narrow Feshbach resonance (\emph{i.e.} ${R^\star \gg |a_{\rm bg}|}$) it is large and negative (${r_{\rm e}^{\rm res}  \sim -2 R^\star}$) with a value which only depends on the general characteristics of the resonance through Eq.~\eqref{eq:Rstar}. 

In the regime where the scattering length is large and positive, the effective range approach supports a shallow bound states with a binding wavenumber given by
\begin{equation}
q_{\rm dim} =\frac{a - \sqrt{a^2 - 2 a r_{\rm e} }}{a r_{\rm e}} .
\label{eq:qdim_effectiverangetrue}
\end{equation}
Equation~\eqref{eq:qdim_effectiverangetrue} and Eq.~\eqref{eq:q_LowE} coincide the each to the other only in the region of asymptotically small detunings ${|{(\mathcal B - \mathcal B_0)} / {\Delta \mathcal B}| \ll 1}$ of narrow resonances (characterized by a large width radius ${R^\star \gg b}$). In this very specific regime, the binding wavenumber of the Feshbach shallow dimer is
\begin{equation}
q_{\rm dim} \simeq \frac{-a +\sqrt{a^2 + 4 R^\star a}}{2R^\star a} 
\label{eq:q_effectiverange}
\end{equation}
which differs from the more general approximate law in Eq.~\eqref{eq:q_narrow}. The effective range approach is thus especially relevant for narrow resonances and for ${|a|\gg |a_{\rm bg}|}$, while for broad resonances it can only supports a qualitative description of the corrections to the universal laws in Eq.~\eqref{eq:Ebp}.

One can also compare Eq.~\eqref{eq:poles} for the  binding wavenumber, with the  equation  written in a similar form but obtained in the framework of the effective range approximation. In the low energy limit where  ${q_{\rm dim}b \ll 1}$ and in the effective range approximation:
\begin{multline}
\frac{q_{\rm dim} a_{\rm bg}}{q_{\rm dim} a_{\rm bg}-1} - \frac{\hbar^2q_{\rm dim}^2}{m{\delta \mu \Delta \mathcal B}} \frac{1}{\left(
1-\frac{ \mathcal B - \mathcal B_0 }{\Delta \mathcal B}\right)\left(1-q_{\rm dim} a_{\rm bg}\right)} \\
 = \frac{\mathcal B - \mathcal B_0} {\Delta \mathcal B} .
\label{eq:bad_interplay}
\end{multline}
Equation~\eqref{eq:bad_interplay} coincides with Eq.~\eqref{eq:poles} only in the regime of asymptotically small detuning {\emph i.e.} ${|{ (\mathcal B - \mathcal B_0)}  / { \Delta \mathcal B}| \ll 1}$) and for ${q_{\rm dim} |a_{\rm bg}|\ll 1}$. Consequently, the interplay between a shape and a Feshbach resonance cannot be properly described by a model relying on an effective range approach. 

\begin{figure}[h]
\includegraphics[width=8cm]{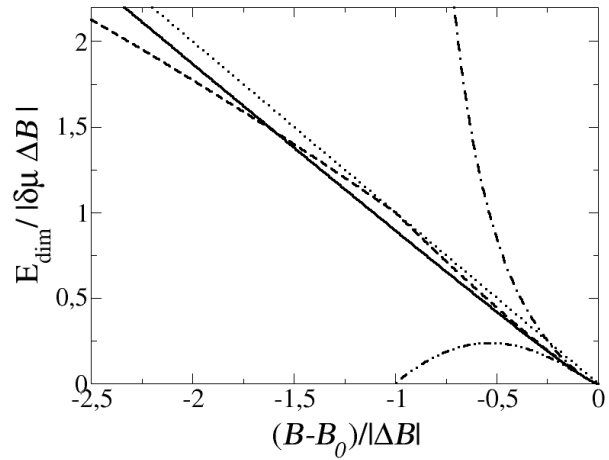}
\caption{Spectrum of the shallow dimer for the narrow resonance of $^{39}$K at 752~G  (${a_{\rm bg}=-35}$~$a_0$, ${\Delta \mathcal B=-0.4}$~G, ${\delta \mu=1.5}$~$\mu_B$, ${R_{\rm vdW}=64.6}$~$a_0$). Solid line: binding energy ${E_{\rm dim}}$ obtained from the two-channel model in units of ${|\delta \mu \Delta \mathcal B|}$ as a function of the relative detuning ${(\mathcal B-\mathcal B_0)/|\Delta \mathcal B|}$; dashed line: approximate law for narrow resonances from Eq.(\ref{eq:q_narrow}). Dotted line: approximation in Eq.~\eqref{eq:intermediate1}, derived in the intermediate regime of narrow resonances; dashed dotted line: binding energy  from Eq.~\eqref{eq:q_effectiverange}; dashed doubly dotted line: binding energy in the effective range approximation \eqref{eq:qdim_effectiverangetrue}.}
\label{fig:K_narrow_dimer}
\end{figure}

In Fig.~(\ref{fig:K_narrow_dimer}) we have plotted the dimer spectrum for the narrow resonance of $^{39}$K at 752~G obtained from the two-channel model. We have added the different approximate laws of Eqs.~(\ref{eq:q_narrow},\ref{eq:intermediate1},\ref{eq:q_effectiverange}). The resonance is sufficiently narrow (${R^\star/|a_{\rm bg}| \sim 34}$) for having a well defined interval of detuning where the intermediate regime of Eq.~\eqref{eq:intermediate_new} holds. 
This figure illustrates the fact that the effective range approximation is relevant only for a small enough relative detuning ${|{(\mathcal B - \mathcal B_0)} /{\Delta \mathcal B}| \ll 1}$) while  the low energy approximation in  Eq.~(\ref{eq:q_narrow}) remains a  good approximation of the exact solution of Eq.~\eqref{eq:poles} of the model even for a finite relative detuning.

\section{Three-body recombination}

\label{sec:three-body}

\subsection{Wave equation for three bosons}

In this subsection, we derive the integral equation which permits to solve the general three-body problem of the present two-channel model. It is thus the first step in order to compute three-body recombination processes. We consider three bosons in their center of mass frame at energy $E$. The state of the system is a coherent superposition of three atoms plus one atom and one molecule:
\begin{multline} 
|\Psi \rangle = \int \frac{d^3K d^3k}{(2\pi)^6} A_{\Kb,\kb} a^\dagger_{\frac{\Kb}{2}+\kb} a^\dagger_{\frac{\Kb}{2}-\kb} 
a^\dagger_{-\Kb}|0\rangle \\
+ \int \dK  \beta_\Kb b^\dagger_{\Kb} a^\dagger_{-\Kb} |0 \rangle .
\label{eq:3b-ansatz}
\end{multline}
In Eq.~\eqref{eq:3b-ansatz}, ${\beta_\Kb}$ (respectively  ${A_{\Kb,\kb}}$) represents after symmetrization the atom-molecule wavefunction (respectively  the wave function for the three atoms in the open channel, where $\Kb$ and $\kb$  are two Jacobi coordinates in momentum space). In what follows, without loss of generality ${A_{\Kb,\kb}}$ is considered to be an even function of $\kb$. 

Projection of the stationary Schr\"odinger's equation ${(E-H)|\Psi\rangle = 0}$, at energy $E$ on the states with one atom and one molecule provides the equation for ${\beta_\Kb}$:
\begin{equation} 
(E_{\rm rel} -E_{\rm mol} ) \beta_\Kb -2 \Lambda \gamma_\Kb = 0
\label{eq:beta}
\end{equation}
where for convenience, we introduced  the function:
\begin{multline}
\gamma_{\Kb} = \int \dk \chi^*_\kb \left(A_{\Kb,\kb} +  2 A_{-\frac{\Kb}{2}+\kb,-\frac{3\Kb}{4}-\frac{\kb}{2}} \right) .
\end{multline}
In Eq.~\eqref{eq:beta}, ${E_{\rm rel}}$ is the relative (or collisional) energy of a pair  of center of mass momentum ${\mathbf K}$:
\begin{equation}
E_{\rm rel}  = E - \frac{3}{4} \frac{\hbar^2 K^2}{m} .
\end{equation}
It represents the energy of a  pair in its center of mass frame while the third particle does not interact with an atom of this pair. The projection of the stationary Schr\"{o}dinger's equation on the states with three atoms gives:
\begin{equation} 
(E_{\rm rel}- 2 \epsilon_{\mathbf k}) A_{\Kb,\kb} -\Lambda \chi_\kb \beta_{\Kb} - g_0 \chi_\kb \gamma_{\Kb} = 0 .
\label{int-eq-A}
\end{equation}
In order to simplify subsequent equations, we introduce the effective wavefunction $\beta^{\rm eff}_\Kb$:
\begin{equation}
\beta^{\rm eff}_\Kb =  \frac{g_0\beta_\Kb}{2\Lambda^2}  
\left[E_{\rm rel} + \delta \mu (\Delta \mathcal B - \mathcal B + \mathcal B_0)\right]
\label{eq:beta_eff}
\end{equation}
where the factor ${\frac{g_0}{2\Lambda^2}}$ can be also written in terms of the molecular energy by using Eq.~\eqref{eq:detuning}. From Eq.~\eqref{eq:beta} we obtain:
\begin{equation} 
(E_{\rm rel} - 2\epsilon_{\mathbf k} ) A_{\Kb,\kb} - \Lambda \chi_\kb \beta^{\rm eff}_\Kb= 0 .
\label{noint-eq-A}
\end{equation}
Similarly to the two-body case [compare with Eqs.~(\ref{eq:intermediaire},\ref{eq:A(k)})], the general solution of Eq.~\eqref{noint-eq-A} can be written as:
\begin{equation} 
A_{\Kb,\kb} = A^{\rm (0)}_{\Kb,\kb} + \frac{\Lambda \chi_\kb \beta^{\rm eff}_\Kb}
{E_{\rm rel} - 2 \epsilon_{\mathbf k}+i0^+}
\label{eq:A}
\end{equation}
where $A^{\rm (0)}$ represents a possible incoming wave of three free atoms, and is thus an eigenstate of the kinetic energy in the center of mass frame. $A^{\rm (0)}$ is non zero only if the energy $E$ is non negative, in this case the second term in the right hand side of Eq.~\eqref{eq:A} represents an outgoing spherical scattered wave, thanks to the ${i0^+}$ term in the denominator. It is clear that for a negative energy, the factor between brackets in Eq.~\eqref{int-eq-A} cannot vanish and the prescription ${E \to E + i0^+}$ can be omitted.

From Eqs.~(\ref{eq:beta},\ref{eq:A}), we deduce a closed integral equation verified by the effective atom-molecule wave function:
\begin{multline}
\frac{m \exp(-m E_{\rm rel} b^2/\hbar^2)}{4 \pi \hbar^2 f(E_{\rm rel})} \beta^{\rm eff}_\Kb+\frac{\gamma^{\rm (0)}_\Kb}{\Lambda} \\
= 
2 \int \dk  \frac{\beta^{\rm eff}_\kb \chi^*_{\frac{\Kb}{2}+\kb}\chi_{\Kb+\frac{\kb}{2}}} {\epsilon_{\mathbf k} + \epsilon_{\mathbf K} + \epsilon_{{\mathbf K} +{\mathbf k}} - E - i 0^+} ,
\label{eq:threebody}
\end{multline}
where, 
\begin{equation} 
\gamma^{\rm (0)}_\Kb = \int \dk \chi^*_\kb \left( A^{\rm (0)}_{\Kb,\kb}
+ 2 A^{\rm (0)}_{-\frac{\Kb}{2} +\kb ,-\frac{3\Kb}{4}- \frac{\kb}{2}} \right).
\label{eq:gamma_0}
\end{equation}
Remarkably this equation has a form similar to the Skorniakov Ter Martirosian (STM) equation \cite{Sko57}. In the limit where ${b \to 0}$, the kernel in the integral of Eq.~\eqref{eq:threebody} coincides with the one of the STM equation. For the more realistic situation where $b$ is finite, the cutoff function ${\chi_{\mathbf k}}$ permits to avoid the Thomas collapse \cite{Tho35}. A remarkable feature of the two-channel separable model used in this paper is that the two-body physics is entirely encapsulated in the diagonal part of the integral equation~\eqref{eq:threebody} only through the two-body scattering amplitude ${f(E)}$. It has been shown recently that this property remains true in the four-body integral equation derived from the same model \cite{Mor11a}.

\subsection{Three-body recombination constant}

We consider the recombination of the three bosons into a Feshbach dimer and one atom in the limit where there is no background dimer. Thus, the following inequalities hold:
\begin{equation}
a > a_{\rm bg} \quad \mbox{and} \quad  a_{\rm bg} < b \sqrt{\pi} .
\end{equation}
The derivation of the recombination rate and recombination constant is obtained along the same lines as in Ref.~\cite{Jon08}. We study the scattering process in the center of mass frame, we thus assume a vanishing total momentum for the incoming wave. Moreover, we  consider the limit of vanishing kinetic energy in the center of mass frame which is relevant for the Bose gas in the degenerate phase. We denote the wavevectors of the three incoming atoms by ${\kb_1^0}$, ${\kb_2^0}$ and ${\kb_3^0}$ and we introduce the total and relative momentum ${\Kb_0=\kb_1^0+\kb_2^0}$ and ${\kb_0=(\kb_1^0-\kb_2^0)/2}$ for the atomic pair 12. The incoming wave function can be written as
\begin{equation}
| \Psi^{\rm (0)} \rangle = \int \frac{d\Kb d\kb}{(2\pi)^6} A^{\rm (0)}_{\Kb,\kb} a^\dagger_{\frac{1}{2}\Kb+\kb}
a^\dagger_{\frac{1}{2}\Kb-\kb}a^\dagger_{-\Kb}|0\rangle ,
\end{equation}
where
\begin{equation}
A^{\rm (0)}_{\Kb,\kb}= (2\pi)^6 \delta(\Kb-\Kb_0)\left[\frac{\delta(\kb-\kb_0)+ \delta(\kb+\kb_0)}{2}\right] .
\end{equation}
In the limit of vanishingly small but positive energy we have:
\begin{equation}
A^{\rm (0)}_{\Kb,\kb} = (2\pi)^6 \delta(\Kb) \delta(\kb) ,
\end{equation}
and we always assume in the following that ${E=0}$. In this limit, the source  term in Eq.~\eqref{eq:threebody} which is given by Eq.~\eqref{eq:gamma_0} is ${\gamma^{\rm (0)}_\Kb = 3(2\pi)^3\delta(\Kb)}$. The atom-molecule wavefunction ${\beta_\Kb}$ can be written as
\begin{multline} 
\beta_\Kb=3 \sqrt{8\pi R^\star} (a-a_{\rm bg}) (2\pi)^3 \delta(\Kb) 
\\ + \sqrt{p_{\rm closed}}\frac{4\pi h(K)}{K^2-K_{\rm dim}^2 -i0^+} .
\label{eq:3b-beta-ansatz}
\end{multline}
The coefficient in front of the delta distribution in Eq.~\eqref{eq:3b-beta-ansatz} is adjusted in order to cancel the source term of the generalized STM equation \eqref{eq:threebody}. In the last term of  Eq.~\eqref{eq:3b-beta-ansatz} we introduced the wave number
\begin{equation}
K_{{\rm dim}} = \frac{2}{\sqrt{3}} q_{{\rm dim}} 
\end{equation}
which corresponds to the relative momentum for the atom and the dimer formed in the inelastic three-body collision. Physically, the  pole at ${K=K_{\rm dim}+i0^+}$ is linked to the outgoing spherical wave for the atom-dimer relative particle. The angular average appearing in Eq.~\eqref{eq:threebody} for the isotropic ansatz of Eq.~\eqref{eq:3b-beta-ansatz}  is performed in Appendix \ref{app:averages}. After calculation one is left with the following inhomogeneous integral equation:
\begin{multline} 
\frac{e^{q_{\rm rel}^2b^2}}{f(E_{\rm rel})} \frac{K h^{\rm eff}(K)}{K^2-K_{\rm dim}^2 -i0^+} 
\\ - \frac{4}{\pi} \int_0^\infty dk\, \frac{k h^{\rm eff}(k)}{k^2-K_{\rm dim}^2 -i0^+}C(K,k,0)e^{-\frac{5}{8}b^2(K^2+k^2)} 
\\ = \frac{1}{\sqrt{p_{\rm closed}}} 
\frac{24 \pi \hbar^2  }{m \Lambda}
\frac{a e^{-\frac{5}{8}b^2 K^2}}{K} ,
\label{eq:tbrec}
\end{multline}
where
\begin{equation}
h^{\rm eff}(K) = \frac{g_0 h(K)}{2\Lambda^2} \left[\delta \mu (\Delta \mathcal B - \mathcal B + \mathcal B_0)-\frac{3}{4}\frac{\hbar^2 K^2}{m}\right] .
\end{equation}
We derive the recombination rate from the solution $h(K)$ by considering formally the three-body system in a large fictitious box of size $L$. We consider the state of three particles in this box with periodic boundary conditions and defined by ${| \psi^{\rm box} \rangle = | \psi \rangle /L^{9/2}}$.  The atom-molecule wavefunction in position space is obtained by a Fourier transform of  Eq.~\eqref{eq:3b-beta-ansatz}:
\begin{equation}
\beta(\rb) = \sqrt{p_{\rm closed}} 4\pi \int \dK  \frac{h(K) e^{i \Kb \cdot\rb}}{K^2-K_{\rm dim}^2 -i 0^+} + \dots
\end{equation}
where ${\rb\equiv \rb_{\rm mol} - \rb_{\rm at}}$ and the dots indicate terms which are not relevant in the recombination process. This provides:
\begin{equation}
\beta(\rb) = \sqrt{p_{\rm closed}} \, h(K_{\rm dim}) \frac{e^{iK_{\rm dim} r}}{r} + \dots
\end{equation}
which corresponds as expected to an outgoing spherical wave for the relative particle composed of the atom and the molecule with the wave number ${K_{\rm dim}}$. In the limit of large separation between the atom and the molecule, the outgoing wave function in the box is:
\begin{equation}
\psi^{\rm box}_{\rm out}(\rb) \simeq \sqrt{p_{\rm closed}} \frac{h(K_{\rm dim})}{L^{9/2}} \frac{e^{iK_{\rm dim} r}}{r} .
\end{equation}
Hence, the probability current for the relative particle of reduced mass $2m/3$ is:
\begin{align}
{\bf j} &= \frac{3\hbar}{4 i m} \left({\psi^{\rm box}_{\rm out}}^* \nabla \psi^{\rm box}_{\rm out} - {\rm c.c.} \right) \nonumber \\
&=\frac{p_{\rm closed}}{L^9} \frac{\hbar K_{\rm dim}}{2m/3}|h(K_{\rm dim})|^2 \frac{\hat{\mathbf e}_{\mathbf r}}{r^2} .
\end{align}
The total flux associated with ${\bf j}$ is obtained by integration over the center of mass position and the full space angle ($4\pi$) for the relative particle:
\begin{equation}
\Phi = 6\pi \frac{p_{\rm closed}}{L^6} \frac{\hbar}{m} |h(K_{\rm dim})|^2 K_{\rm dim} .
\end{equation}
The outgoing dimer has a probability ${p_{\rm closed}}$ to be in the closed  channel, we thus divide this total flux by the corresponding probability in order to get the probability flux of dimer formation. The atom loss rate is then
\begin{equation}
\frac{d{N}_{\rm dim}^{\rm box}}{dt} = -\frac{\Phi}{p_{\rm closed}}= -  \frac{6\pi\hbar }{m} K_{\rm dim} |h(K_{\rm dim})|^2 \frac{1}{L^6} .
\label{eq:ndot-3b}
\end{equation}
The general definition of  the three-body recombination constant ${\alpha_{\rm rec}}$ for a system of $N$ particles is:
\begin{equation} 
\frac{d{N}_{\rm dim}^{\rm box}}{dt} = -\alpha_{\rm rec} \frac{N(N-1)(N-2)}{L^6} .
\label{eq:ndot-def}
\end{equation}
In the present case ${N=3}$ in Eq.~\eqref{eq:ndot-def} and by identification with Eq.~\eqref{eq:ndot-3b} we obtain the link between the recombination constant and the  atom-molecule wave function:
\begin{equation} 
\alpha_{\rm rec} =  \frac{\pi \hbar }{m} K_{\rm dim}|h(K_{\rm dim})|^2 .
\label{eq:a_rec}
\end{equation}
Finally, the determination of the recombination constant reduces to the calculation of $h(K_{\rm dim})$, that is to the calculation of the residue of the atom-molecule wavefunction at the wave number ${K=K_{\rm dim}}$. 

We now apply Eq.~\eqref{eq:a_rec} to  the case of narrow resonances  in the regime  ${R^{\star}\gg |a|}$  where the integral term in Eq.~\eqref{eq:tbrec} is negligible for asymptotically large values of $R^\star$. We evaluate ${h^{\rm eff}(K_{\rm dim})}$ to the lowest order by neglecting the integral term, thereby obtaining
\begin{equation}
h^{\rm eff}(K) \simeq \frac{1}{\sqrt{p_{\rm closed}}}  \frac{24 \pi \hbar^2  }{m \Lambda} \frac{a}{K^2} f(E_{\rm rel})(K^2-K_{\rm dim}^2) .
\end{equation}
Large values of the width radius $R^\star$ are reached in the intermediary regime of Eq.~\eqref{eq:intermediate_new}, so that we use the approximate law Eq.~\eqref{eq:intermediate1} for the dimer's binding energy. Moreover, we use the approximation ${p_{\rm closed}\simeq 1}$ and we find the dominant contribution:
\begin{equation}
h^{\rm eff}(K_{\rm dim}) \simeq \frac{24\pi \hbar^2}{m\Lambda} a(a-a_{\rm bg}) .
\end{equation}
Finally, in the intermediate regime the recombination constant can be approximated by the law:
\begin{equation}
\alpha_{\rm rec} \operatornamewithlimits{=}_{R^{\star} \gg |a| } \frac{192 \pi^2 a^2 \hbar}{m} \sqrt{3 R^\star  \left(a-a_{\rm bg}\right)^3}  + \dots
\label{eq:a_rec_narrow}
\end{equation} 
\begin{figure}[h]
\includegraphics[width=8cm]{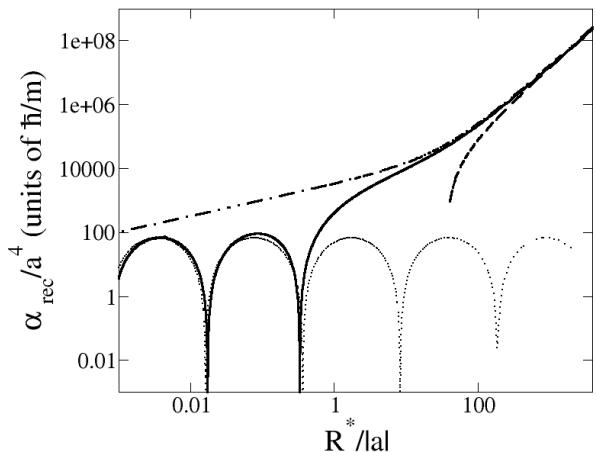}
\caption{Three-body recombination constant $\alpha_{\rm rec}$ in units of $\hbar/m$ for the narrow resonance in $^{39}K$ in the vicinity of  752 G. Solid line: numerical solution of Eq.~\eqref{eq:tbrec} for positive value of $a$; dashed line: numerical solution of Eq.~\eqref{eq:tbrec} for ${a_{\rm bg}<a<0}$; dotted line: analytical solution of Ref.~\cite{Gog08} valid for large values of the scattering length; dashed-dotted line:  approximate law of Eq.~\eqref{eq:a_rec_narrow} valid for large values of the ratio $R^\star/|a|$. In order to guide the eyes, this last curve is also plotted in the region where ${R^\star/|a|}$ is small.}
\label{fig:K_narrow_tbrec}
\end{figure}
This is the main result of this paper which is a generalization of Eq.~(16) in Ref.~\cite{Pet04b} obtained in the effective range approximation and where the supplementary condition ${a \gg a_{\rm bg}}$ was assumed. The deviation of Eq.~\eqref{eq:a_rec_narrow} from this last law is substantial in the case where the background scattering length is negative and the ratio $R^\star/|a|$ can be arbitrarily small for a fixed value of $R^\star$ while a shallow dimer still exists. In this regime, the recombination constant tends to zero with an anomalous quadratic law in the scattering length $a$ (the usual law brought by dimensional analysis is in $a^4$). This regime can be reached for instance in the case of the narrow resonance in potassium near 752~Gauss (see Fig.~\ref{fig:K_narrow_dimer}). We illustrate this example in Fig.~\eqref{fig:K_narrow_tbrec}, where the asymptotic law in Eq.~\eqref{eq:a_rec_narrow} is compared to the exact numerical calculation. In the region where the scattering length $a$ is large (and positive) with respect to the width radius $R^\star$, we recover the oscillatory law in log-scale which is characteristics of the Efimov effect and was derived analytically in Ref.~\cite{Gog08} in the framework of a two-channel model strictly equivalent to the effective range approximation.
\begin{figure}[h]
\includegraphics[width=8cm]{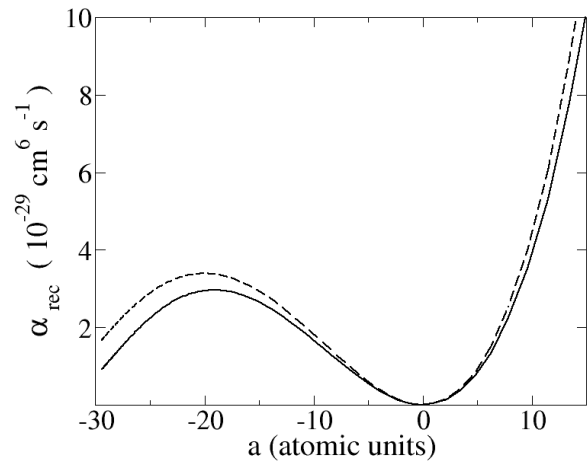}
\caption{Three-body recombination constant $\alpha_{\rm rec}$ in the region of small scattering length (${|a|<|a_{\rm bg}|}$) for the narrow resonance in $^{39}K$ in the vicinity of  752 G. Solid line: numerical solution of Eq.~\eqref{eq:tbrec}; dashed line: approximate law of Eq.~\eqref{eq:a_rec_narrow} valid for large values of the ratio $R^\star/|a|$.}
\label{fig:K_narrow_crossing}
\end{figure}
In Fig.~\eqref{fig:K_narrow_crossing}, we plotted the three-body recombination constant for the same species in the region of zero crossing for the scattering length ${(\mathcal B=\mathcal B_0 + \Delta \mathcal B)}$ where it vanishes quadratically.

\section{Conclusions}

In this paper, we used a separable two-channel model for the description of the magnetic Feshbach resonance in ultracold atoms. We studied the deviation of the dimer's binding energy with respect to the general threshold law Eq.~\eqref{eq:Ebp} in the different regimes that can be achieved experimentally. For narrow resonances, we defined a intermediate regime of detuning where the background scattering length is a relevant parameter. In this regime, we derived approximate laws for the binding energy of the shallow dimer and for the three-body recombination rate. These results show the interest of using a two-channel model for the description of ultracold gases in the resonant regime. In particular it appears as an useful tool for providing universal properties (that is, properties which are not explicitly dependent of the short range details of the interactions) in regimes where the effective range approximation is not quantitative. 

{\bf Acknowledgments}-- Y. Castin and M. Zaccanti are acknowledged for thorough discussions on the subject. A. Simoni is acknowledged for providing us the numerical data of Fig.~(\ref{fig:compardimers1}). LPTMC is UMR 7600 of CNRS and its Cold Atoms group is associated with IFRAF.

\appendix

\section{Angular averages}

\label{app:averages}

The ansatz for $\beta_\Kb$ used in this paper is isotropic. Thus in the homogeneous part of Eq.~\eqref{eq:threebody} one has to perform the following angular average:
\begin{equation}
I(\Kb, k,q) = \int \frac{d\Omega_{\hat \kb}}{4\pi}\, \frac{e^{-b^2 \Kb \cdot \kb}}
{q^2+K^2+k^2+\Kb \cdot \kb}
\end{equation}
(in the recombination process considered in the present paper, $q=0$). To perform the angular integration we use spherical polar coordinates of polar axis $\Kb/K$. After the integration over the azimutal $\varphi$ angle, we are left with
\begin{equation}
I(\Kb, k,q) = \frac{C(K,k,q)}{Kk} \label{eq:I} .
\end{equation}
In Eq.(\ref{eq:I}), C(K,k,q) is defined by:
\begin{equation}
C=\int_{-1}^1 \frac{dx}{2}\,\frac{e^{-tx}}{v+x}=\frac{e^{vt}}{2} \left[
E_1(vt-t)-E_1(vt+t)\right]
\end{equation}
where ${E_1(z) = \int_1^{\infty} ds\, \frac{e^{-sz}}{s}}$ is the exponential integral, and the variables $(v,t)$ are:
\begin{equation}
v = \frac{q^2+K^2+k^2}{Kk} \qquad \qquad t=b^2 Kk .
\end{equation}
In the zero-range limit $b \to 0$ we obtain the simple result:
\begin{equation}
C = \frac{1}{2} \mbox{ ln}\left(\frac{q^2+K^2+k^2+Kk}{q^2+K^2+k^2-Kk}\right)
\end{equation}
which coincides with the result of Skorniakov and Ter Martirosian \cite{Sko57}.

%\end{multicols}


\begin{thebibliography}{0}
\expandafter\ifx\csname natexlab\endcsname\relax\def\natexlab#1{#1}\fi
\expandafter\ifx\csname bibnamefont\endcsname\relax
  \def\bibnamefont#1{#1}\fi
\expandafter\ifx\csname bibfnamefont\endcsname\relax
  \def\bibfnamefont#1{#1}\fi
\expandafter\ifx\csname citenamefont\endcsname\relax
  \def\citenamefont#1{#1}\fi
\expandafter\ifx\csname url\endcsname\relax
  \def\url#1{\texttt{#1}}\fi
\expandafter\ifx\csname urlprefix\endcsname\relax\def\urlprefix{URL }\fi
\providecommand{\bibinfo}[2]{#2}
\providecommand{\eprint}[2][]{\url{#2}}

\end{thebibliography}


\begin{thebibliography}{99}

%Unified Theory of Nuclear Reactions
\bibitem{Fes58} H. Feshbach, 
%\href{http://dx.doi.org/10.1016/0003-4916(58)90007-1}
{Ann. Phys. (N. Y.) {\bf 5}, 357 (1958)}; 
%doi:10.1016/0003-4916(58)90007-1
{\sl ibid.} 
%\href{http://dx.doi.org/10.1016/0003-4916(62)90221-X}
{{\bf 19}, 287 (1962)}.
%doi:10.1016/0003-4916(62)90221-X

%Observation of Feshbach resonances in a Bose~Einstein condensate -> resonance etroite Na23
\bibitem{Ino98} S.~Inouye, M.R.~Andrews, J.~Stenger, H.-J.~Miesner, D.M.~Stamper-Kurn, and W.~Ketterle, 
%\href{http://www.nature.com/nature/journal/v392/n6672/abs/392151a0.html}
{Nature (London) {\bf 392}, 151 (1998)}.
%doi:10.1038/32354

%Evidence for Efimov quantum states in an ultracold gas of caesium atoms; Accepted 2 February 2006
\bibitem{Kra06} T. Kraemer, M. Mark, P. Waldburger, J. G. Danzl, C. Chin, B. Engeser, A. D. Lange, 
K. Pilch, A. Jaakkola, H.-C. N\"{a}gerl and R. Grimm, 
%\href{http://www.nature.com/nature/journal/v440/n7082/abs/nature04626.html}
{Nature {\bf 440}, 315 (2006)}.
%doi:10.1038/nature04626

%Observation of an Efimov-like resonance in ultracold atom-dimer scattering
\bibitem{Kno09} S. Knoop, F. Ferlaino, M. Mark, M. Berninger, H. Schoebel, H.-C. Naegerl, R. Grimm, 
%\href{http://www.nature.com/nphys/journal/v5/n3/abs/nphys1203.html}
{Nature Physics {\bf 5}, 227 (2009)}.
%doi:10.1038/nphys1203

%Observation of an Efimov spectrum in an atomic system
\bibitem{Zac09} M. Zaccanti, B. Deissler, C. D'Errico, M. Fattori, M. Jona-Lasinio, S. M\"{u}ller, G. Roati, M. Inguscio, G. Modugno, 
%\href{http://www.nature.com/nphys/journal/v5/n8/abs/nphys1334.html}
{Nature Phys. {\bf 5}, 586 (2009)}.

%Evidence for Universal Four-Body States Tied to an Efimov Trimer
\bibitem{Fer09} F. Ferlaino, S. Knoop, M. Berninger, W. Harm, J. P. D'Incao, H.-C. N\"{a}gerl, R. Grimm, 
%\href{http://link.aps.org/doi/10.1103/PhysRevLett.102.140401}
{Phys. Rev. Lett. {\bf 102}, 140401 (2009)}.
%URL:http://link.aps.org/doi/10.1103/PhysRevLett.102.140401
%DOI:10.1103/PhysRevLett.102.140401

%Universality in Three- and Four-Body Bound States of Ultracold Atoms
\bibitem{Pol09} S.E. Pollack, D. Dries, R.G. Hulet, 
%\href{http://www.sciencemag.org/cgi/content/abstract/sci;326/5960/1683}
{Science {\bf 326}, 1683 (2009)}.
%DOI: 10.1126/science.1182840

%Mesure des pertes a 3-corps en fonction de B
%Strongly Enhanced Inelastic Collisions in a Bose-Einstein Condensate near Feshbach Resonances suite de l'article dans Nature
\bibitem{Ste99} J. Stenger, S. Inouye, M.R. Andrews, H.-J. Miesner, D.M. Stamper-Kurn, and W. Ketterle, 
%\href{http://link.aps.org/doi/10.1103/PhysRevLett.82.2422}
{Phys. Rev. Lett. {\bf 82}, 2422 (1999)}.
%URL:http://link.aps.org/doi/10.1103/PhysRevLett.82.2422
%DOI:10.1103/PhysRevLett.82.2422

%Resonances in ultracold collisions of 6Li, 7Li, and 23Na 
\bibitem{Moe95} A. J. Moerdijk, B. J. Verhaar, and A. Axelsson, 
%\href{http://link.aps.org/doi/10.1103/PhysRevA.51.4852}
{Phys. Rev. A {\bf 51}, 4852 (1995)}.
%URL:http://link.aps.org/doi/10.1103/PhysRevA.51.4852
%DOI:10.1103/PhysRevA.51.4852

%Calculation of the scattering length in atomic collisions using semiclassical approximation
\bibitem{Gri93} G.F. Gribakin and V.V. Flambaum, 
%\href{http://link.aps.org/doi/10.1103/PhysRevA.48.546}
{Phys. Rev. A {\bf 48}, 546 (1993)}.
%URL:http://link.aps.org/doi/10.1103/PhysRevA.48.546
%DOI:10.1103/PhysRevA.48.546

%Collisional dynamics of ultra-cold atomic gases.
\bibitem{Dal98} J. Dalibard, Proceedings of the International School of  Physics 'Enrico  Fermi', Course CXL: 
{\sl 'Bose-Einstein condensation in gases'}, Varenna 1998,  M. Inguscio, S. Stringari, C. Wieman edts.

%Excited Thomas-Efimov levels in ultracold gases
\bibitem{Lee07} M.D. Lee, T. K\"{o}hler and P.S. Julienne, 
%\href{http://link.aps.org/doi/10.1103/PhysRevA.76.012720}
{Phys. Rev. A {\bf 76}, 012720 (2007)}. 
%URL:http://link.aps.org/doi/10.1103/PhysRevA.76.012720
%DOI:10.1103/PhysRevA.76.012720


%Feshbach resonances in ultracold gases
\bibitem{Chi10} C. Chin, R. Grimm, P. Julienne, and E. Tiesinga, 
%\href{http://link.aps.org/doi/10.1103/RevModPhys.82.1225}
{Rev. Mod. Phys. {\bf 82}, 1225 (2010)}.
%URL:http://link.aps.org/doi/10.1103/RevModPhys.82.1225
%DOI:10.1103/RevModPhys.82.1225

%Number of closed-channel molecules in the BEC-BCS crossover
\bibitem{Wer09} F. Werner, L. Tarruell, and Y. Castin,
%\href{http://www.springerlink.com/index/84j378181w489491.pdf}
{Eur. Phys. J. B {\bf 68}, 401 (2009)}.
%DOI:10.1140/epjb/e2009-00040-8

%Three Resonant Ultra-Cold Bosons: Off-Resonance Effects
\bibitem{Jon10} M. Jona-Lasinio, L. Pricoupenko, Phys. Rev. Lett. {\bf 104}, 023201 (2010).

%{\sl'Integral equations for the four-body problem'};
\bibitem{Mor11a} C. Mora, Y. Castin, L. Pricoupenko, Comptes Rendus Physique {\bf 12}, 71 (2011).

%The Efimov effect in lithium 6
\bibitem{Nai10} P. Naidon and M. Ueda, Comptes Rendus Physique, {\bf 12}, 13 (2011).

%Three fully polarized fermions close to a p-wave Feshbach resonance
\bibitem{Jon08}  M. Jona-Lasinio, L. Pricoupenko and Y. Castin, Phys. Rev. A {\bf 77}, 043611 (2008).

%Feshbach resonances in ultracold 39K
\bibitem{Der07} C. D'Errico, M. Zaccanti, M. Fattori, G. Roati, M. Inguscio, G. Modugno, A. Simoni, 
New Jour. Phys. {\bf 9}, 223 (2007).

%Three-Boson Problem near a Narrow Feshbach Resonance
\bibitem{Pet04b} D.S. Petrov, 
%\href{http://link.aps.org/doi/10.1103/PhysRevLett.93.143201}
{Phys. Rev. Lett. {\bf 93}, 143201 (2004)}.
%URL:http://link.aps.org/doi/10.1103/PhysRevLett.93.143201
%DOI:10.1103/PhysRevLett.93.143201

%Isotropic contact forces in arbitrary representation: Heterogeneous few-body problems and low dimensions
\bibitem{Pri11} L. Pricoupenko, Phys. Rev. A {\bf 83}, 062711 (2011).
%URL: http://link.aps.org/doi/10.1103/PhysRevA.83.062711
%DOI: 10.1103/PhysRevA.83.062711

\bibitem{Lan99} L. Landau and E. Lifchitz, {\sl 'Quantum Mechanics'} (Butterworth-Heinemann, Oxford, 1999).

\bibitem{Tay72} J.R. Taylor,  {\sl'Scattering theory'}, Wiley, New York (1972).

%Interaction n-p for the'Diplon' ie H2
\bibitem{Bet35} H. Bethe and R. Peierls, 
%\href{http://dx.doi.org/10.1098/rspa.1935.0010}
{Proc. R. Soc. London, Ser. {\bf A} {\bf 148}, 146 (1935)}.

%Analytical solution of the bosonic three-body problem
\bibitem{Gog08} A.O. Gogolin, C. Mora, R. Egger, 
%\href{http://link.aps.org/doi/10.1103/PhysRevLett.100.140404}
{Phys. Rev. Lett. {\bf 100}, 140404 (2008)}.
%URL:http://link.aps.org/doi/10.1103/PhysRevLett.100.140404
%DOI:10.1103/PhysRevLett.100.140404

%Crossover in the Efimov spectrum
\bibitem{Pri10b} L. Pricoupenko, Phys. Rev. A {\bf 82}, 043633 (2010). 

%Production of cold molecules via magnetically tunable Feshbach resonances
\bibitem{Koh06} Thorsten K\"{o}hler, Krzysztof G\'{o}ral, and Paul S. Julienne, 
%\href{http://link.aps.org/doi/10.1103/RevModPhys.78.1311}
{Rev. Mod. Phys. {\bf 78}, 1311 (2006)}.

%Feshbach resonances with large background scattering length: Interplay with open-channel resonances,
\bibitem{Mar04} B. Marcelis, E. G. M. van Kempen, B. J. Verhaar, and S. J. J. M. F. Kokkelmans, Phys. Rev. A {\bf 70}, 012701 (2004).

%Signatures of universal four-body phenomena and their relation to the Efimov effect
\bibitem{Ste09} Javier von Stecher, Jose P. D'Incao, Chris H. Greene, 
%\href{http://www.nature.com/nphys/journal/v5/n6/abs/nphys1253.html}
{Nature Physics {\bf 5}, 417 (2009)}.
%doi:10.1038/nphys1253

%Spectroscopy of ultracold trapped cesium Feshbach molecules
\bibitem{Mar07} M. Mark, F. Ferlaino, S. Knoop, J. G. Danzl, T. Kraemer, C. Chin, H.-C. N\"{a}gerl, and R. Grimm,  
%\href{htp://link.aps.org/doi/10.1103/PhysRevA.76.042514}
{Phys. Rev. A {\bf 76}, 042514 (2007)}.
%URL:http://link.aps.org/doi/10.1103/PhysRevA.76.042514
%DOI:10.1103/PhysRevA.76.042514

\bibitem{Sko57} G.V. Skorniakov and K.A. Ter-Martirosian, Sov. Phys. JETP {\bf 4}, 648 (1957).

%The Interaction Between a Neutron and a Proton and the Structure of H3
\bibitem{Tho35} L.H. Thomas,
%\href{http://link.aps.org/doi/10.1103/PhysRev.47.903}
{Phys. Rev. {\bf 47}, 903 (1935)}.
%URL:http://link.aps.org/doi/10.1103/PhysRev.47.903
%DOI:10.1103/PhysRev.47.903


\end{thebibliography}
\end{document}